\documentstyle[12pt]{article}
\begin{document}
%======================================%
%<<<<<<<<<<<  DEFINITION  >>>>>>>>>>>>>%
%======================================%
\newcommand{\gsim}{\mbox{\raisebox{-1.0ex}{$\stackrel{\textstyle >}
{\textstyle \sim}$ }}}
\newcommand{\lsim}{\mbox{\raisebox{-1.0ex}{$\stackrel{\textstyle <}
{\textstyle \sim}$ }}}
\newcommand{\bfx}{{\bf x}}
\newcommand{\bfy}{{\bf y}}
\newcommand{\bfr}{{\bf r}}
\newcommand{\bfk}{{\bf k}}
\newcommand{\bkp}{{\bf k'}}
\newcommand{\order}{{\cal O}}
\newcommand{\beq}{\begin{equation}}
\newcommand{\eeq}{\end{equation}}
\newcommand{\beqa}{\begin{eqnarray}}
\newcommand{\eeqa}{\end{eqnarray}}
\newcommand{\lmk}{\left(}
\newcommand{\rmk}{\right)}
\newcommand{\lkk}{\left[}
\newcommand{\rkk}{\right]}
\newcommand{\lnk}{\left\{}
\newcommand{\rnk}{\right\}}
\newcommand{\call}{{\cal L}}
\newcommand{\calh}{{\cal H}}
\newcommand{\ppp}{\partial}
\newcommand{\dq}{\frac{dq}{d\tau}}
\newcommand{\ddq}{\frac{d^2q}{d\tau^2}}
\newcommand{\dddq}{\frac{d^3q}{d\tau^3}}
\newcommand{\ddddq}{\frac{d^4q}{d\tau^4}}
\thispagestyle{empty}
%======================================%
%<<<<<<<<<<<< TITLE PAGE >>>>>>>>>>>>>>%
%======================================%
\thispagestyle{empty}
{\baselineskip0pt
\leftline{\large\baselineskip16pt\sl\vbox to0pt{\hbox{Department of Physics} 
               \hbox{The University of Tokyo}\vss}}
\rightline{\large\baselineskip16pt\rm\vbox to20pt
{\hbox{UTAP-231, RESCEU-17/96}
           \hbox{OCHA-PP-65}
               \hbox{\today}
\vss}}%
}
\vskip15mm
\begin{center}
{\large\bf Small Thermal Fluctuations on a Large Domain}
\end{center}
\begin{center}
{\large Tetsuya Shiromizu} \\
\sl{Department of Physics, The University of Tokyo, Tokyo 113, Japan \\
and \\
Research Center for the Early Universe(RESCEU), \\ The University of Tokyo, 
Tokyo 113, Japan}
\end{center}
\begin{center}
{\large Masahiro Morikawa} \\
\sl{Department of Physics, Ochanomizu University, Tokyo 112, Japan}
\end{center}

%\begin{center}
%{\it Submitted to Physical Review Letter}
%\end{center}
%======================================%
%<<<<<<<<<<<<< ABSTRACT >>>>>>>>>>>>>>>% 
%======================================%
\begin{abstract} 
Weak first-order phase transitions proceed with percolation of new phase. 
The kinematics of this process is clarified from the point of view of subcritical 
bubbles.  We examine the effect of small subcritical bubbles around a large 
domain of asymmetric phase by introducing an effective geometry. The percolation 
process can be understood as a perpetual growth of the large domain aided by the 
small subcritical bubbles.  
\end{abstract}
%\vfill
\vskip1cm
%\multicols{2}
%======================================%
%<<<<<<<<<<<< SECTION I  >>>>>>>>>>>>>>%
%======================================%
%\baselineskip25pt
\par
Recently crucial effects of small rapid fluctuations 
-subcritical bubbles\cite{gl} are actively investigated for the electroweak 
baryogenesis\cite{ctm} and for the inflationary cosmology. 
We could previously show, by studying the effect of subcritical bubbles, that 
the phase mixing is achieved in weakly first order phase transitions.  
However the global achievement of the phase transition -percolation- has not yet 
been explained from the point of view of subcritical bubbles.  
The difficulty lies on the previous treatment of subcritical bubbles; one always 
assumed spherically symmetry for the configuration of subcritical bubbles and 
interactions among them were neglected. 
However, it is possible to consider these effects easily within the point of 
view of subcritical bubbles.  After the phase mixing is attained, there 
appear large and stable domains of asymmetric phase. If this domain were 
isolated in thermal fluctuations, then this domain would eventually collapse due 
to the surface tension of itself.  However if we properly consider the effect of 
small fluctuations of subcritical bubbles around the large domain, this domain 
is stabilized against the collapse and eventually grows.  This is the mechanism 
which we would like to clarify in this paper.  This growth can be seen as the 
process of percolation of the system.  Our study has been inspired by the 
interesting work by Gleiser et al.\cite{new}.  
However in our case, both the collapse and growth of the domain are 
automatically taken into account and we worry about the 
introducing an extra term 
which guarantees the decay as claimed in \cite{new}.  
\par
In this letter, we first show that a large asymmetric domain is stabilized 
by subcritical bubbles around the wall. 
For simplicity, we assume that 
the shape of the large domain is spherical with 
radius $R(t)$ and volume $V_+(t)$. 
Hereafter we simply call the subcritical bubbles as bubbles. The time variation 
of the volume $V_+(t)$ is composed from the kinematical and subcritical bubble 
contributions% 
%%%begin footnote%%%%%%%%%%%%%%%%%%%%%%%
\footnote{%
As we stated Gleiser et al. have considered that a domain without bubbles always collapses; 
in place of the first term on the right-hand side of this equation, they put a 
term $-4 \pi R^2v$ where $v$ is the wall velocity of 
the collapsing domain. We suspect this treatment; actual thermal 
fluctuations around a domain have both collapsing and expanding 
effects on the domain. Furthermore, they compare their results with  computer 
simulations\cite{simu}.  In this simulation the 
``final" fraction($\gamma_{\rm f}$) 
of the symmetric phase depends on the strength of the self-coupling $\lambda$ of 
the Higgs field which determines the mass. 
For sufficiently large $\lambda$, this fraction becomes $0.5$ within the Hubble 
time.  However for smaller values of $\lambda$, it never reach $0.5$ .  In this 
latter case, the system is always in non-equilibrium state.  
On the other hand, they compare the {\it thermal equilibrium} value estimated 
through the Boltzmann equation with the above {\it non-equilibrium} value. The 
Boltzmann equation they use is given by
%============< EQUATION >==============%
%
\begin{equation}
\partial_t F(\phi, R) \simeq -v\partial_RF(\phi, R)+(1-2f)G(\phi, R),
%\label{bolt}
\end{equation}
%
%======================================%
where $F$ and $G$ are the number density  and 
the creation rate of the subcritical bubble inside the range 
$(\phi \sim \phi + \delta \phi, R \sim R +\delta R)$, respectively. 
In the usual case, the first term of the right-hand side in the above equation 
vanishes and the equilibrium value should be $\gamma_{\rm eq} =1/2$. Of course 
the result in the simulation is $\gamma^{\rm simul.}=1/2$ if one 
continues the numerical calculation sufficiently beyond the Hubble time.     }.  
%%%end footnote%%%%%%%%%%%%%%%%%%%%%%%
;
%============< EQUATION >==============%
%
\begin{equation}
\frac{dV_+(t)}{dt}=4  \pi R^2 \frac{dR}{dt}+(\Gamma_+ 
\Delta V) \frac{4 \pi}{3} \langle R \rangle_+^3 -
(\Gamma_0 \Delta V') \frac{4 \pi}{3} \langle R \rangle_0^3,
\label{vol}
\end{equation}
%
%======================================%
where only the contribution from bubbles near wall of 
the large domain is included in the second and third terms in the 
right-hand-side of this equation. 
$\Gamma_+$ and $\Gamma_0$ are the creation rates per unit volume of a bubble and 
an anti-bubble, respectively: 
%============< EQUATION >==============%
%
\begin{equation}
\Gamma_+ \simeq m_+^4(T) {\rm exp}\lkk -\beta F_+( \langle R 
\rangle_+) \rkk
\end{equation}
%
%======================================%
and
%============< EQUATION >==============%
%
\begin{equation}
\Gamma_0 \simeq  m_0^4(T) {\rm exp}\lkk -\beta F_0( \langle R 
\rangle_0) \rkk, 
\end{equation}
%
%======================================%
where $F_{0, +}$ and $\langle R \rangle_{0, +}$ are the free energy and  
the averaged size of a bubble and an anti-bubble, respectively. 
As $F_0 \simeq F_+$, $m_0(T) \simeq m_+(T)$ and 
$\langle R \rangle_+ \simeq \langle R \rangle_0 :=\langle R \rangle$ 
near $T=T_c$ at which two vacua degenerate, approximately $\Gamma_+=\Gamma_0=:\Gamma$ holds and therefore
%============< EQUATION >==============%
%
\begin{equation}
\frac{dV_+(t)}{dt}=4  \pi R^2 \frac{dR}{dt}+\Gamma  
\lmk \Delta V - \Delta V'\rmk \frac{4 \pi}{3} \langle R \rangle^3.
\end{equation}
%
%======================================%
Furthermore, as $\Delta V -\Delta V' \simeq 32 \pi R \langle R \rangle^2$, this 
equation reduces to
%============< EQUATION >==============%
%
\begin{equation}
\frac{dV_+(t)}{dt}=4  \pi R^2 \frac{dR}{dt}+ \frac{128 \pi^2}{3} 
\Gamma R \langle R \rangle^5.
\label{effvol1}
\end{equation}
%
%======================================%
This is the relation between the spherical symmetric volume and its radius 
modified by the bubbles around the large domain.  
A cute and elegant way to express this important relation is to introduce a 
fictitious geometry of non-Euclidean space whose metric is given by 
%============< EQUATION >==============%
%
\begin{equation}
d \ell^2=dr^2+a(t)r^2d \Omega^2_2.
\label{effgeo}
\end{equation}
%
%======================================%
Deviation of the ``scale factor'' $a$ from 1 represents the non-trivial effect 
from the last term in eq.(\ref{effvol1}).  
The time variation of the volume in this geometry (\ref{effgeo}) becomes
%============< EQUATION >==============%
%
\begin{equation}
\frac{dV_+}{dt} = \frac{d}{dt}\lmk  4 \pi \int^{R(t)}_0 drr^2a(t) \rmk=
\frac{d}{dt}\lmk \frac{4 \pi}{3}R^3(t)a(t) \rmk.
\label{effvol2}
\end{equation}
%
%======================================%
Equating eq.(\ref{effvol1}) and eq.(\ref{effvol2}), one obtains 
an `Einstein' equation 
for a `scale' factor $a(t)$;
%============< EQUATION >==============%
%
\begin{equation}
\frac{d}{dt}\lmk \frac{4 \pi}{3}R^3(t)[a(t)-1] \rmk = 
\frac{128 \pi^2}{3} \Gamma R \langle R \rangle^5.
\end{equation}
%
%======================================%
The solution of this equation is 
%============< EQUATION >==============%
%
\begin{equation}
a(t)=1+\frac{32 \pi \Gamma \langle R \rangle^5}{R^3(t)} 
\int^t_0dt'R(t') =1+ \gamma x^{-3}(\tau) 
\int^\tau_0 d\tau' x(\tau'),
\end{equation}
%
%======================================%
where $\gamma:=32 \pi \Gamma \langle R \rangle^4$, $x:=
R/\langle R \rangle $ and $\tau:=t/\langle R \rangle$. 
\par
Next, we construct an equation of motion for `matter'($R(t)$). 
It is obvious that its Lagrangian is given by 
%============< EQUATION >==============%
%
\begin{eqnarray}
L & = & \int d^3x {\sqrt q}{\cal L} \nonumber \\
  & = & 4 \pi a(t) \int^\infty_0 drr^2 {\cal L} \nonumber \\
  & = & a(t)L_0,
\end{eqnarray}
%
%======================================%
where
%============< EQUATION >==============%
%
\begin{equation}
L_0=\frac{1}{2}M(R)\lmk \frac{dR}{dt} \rmk^2-V_0(R)
\end{equation}
%
%======================================%
and $M(R):=\frac{15{\pi}^{3/2}{\phi}_+^2R}{8{\sqrt {2}}}$. 
Here we have used the form $\phi(\bfx)=\phi_+{\rm exp}[-|\bfx|^2/R^2(t)]$ for 
the large domain as previously\cite{sb} \footnote{
For simplicity, we assumed that the configuration 
is Gaussian, but this is true if the domain has the thick wall. 
In general, the wall can be thin. However we expect that the non-Gaussian 
extension cannot give the drastic change on the present qualitative result.} . 
Near the critical temperature the potential becomes 
%============< EQUATION >==============%
%
\begin{equation}
V(R) = a(t)V_0(R) \simeq  \frac{2}{5}a(t)M(R). 
\end{equation}
%
%======================================%
Let us define a new variable:$q(\tau):=\int^\tau_0d\tau' 
x(\tau')$. Then the Lagrangian becomes 
%============< EQUATION >==============%
%
\begin{eqnarray}
{\cal L}\lmk q,\frac{dq}{d\tau}, \frac{d^2q}{d\tau^2} \rmk & := & 
\frac{1}{M(\langle R \rangle )} L \nonumber \\
\nonumber \\ & = & 
\lkk 1+\gamma \lmk \frac{dq}{d\tau} \rmk^{-3} q \rkk 
\lkk \frac{1}{2} \frac{dq}{d \tau}  \lmk \frac{d^2q}{d \tau^2} \rmk^2 
-\frac{2}{5} \frac{dq}{d\tau} \rkk.
\end{eqnarray}
%
%======================================%
The corresponding Euler-Lagrange equation becomes 
%============< EQUATION >==============%
%
\begin{eqnarray}
10 \lmk \dq \rmk^5 \ddddq & + & 10 \gamma q  \lmk \ddq \rmk^2 \ddddq 
 +  20 \lmk \dq \rmk^4 \ddq \dddq \nonumber \\ 
& - & 40 \gamma q \dq \ddq \dddq + 
20 \gamma \lmk \dq \rmk^3 \dddq \nonumber \\ 
& + & 30 \gamma q \lmk \ddq \rmk^3 
-15 \gamma \lmk \dq \rmk^2 \lmk \ddq \rmk^2 \nonumber \\
& + & 24 \gamma q \ddq 
-12 \gamma \lmk \dq \rmk^2 =0.
\end{eqnarray}
%
%======================================%
It is hopeless to solve this equation analytically; numerical calculation is 
necessary.  
The results of numerical calculations are depicted in Fig. 1 and 2. 
They show the time evolution of $x$ the radius $R$ normalized by 
$\langle R \rangle$.  
In Fig. 1, we changed $\gamma$ fixing the initial velocity ${\dot 
x}(0)$. 
In Fig. 2, we fixed $\gamma$ and changed the initial velocity. 
>From the figures one easily see that the domain always bounces 
by the fluctuation of bubbles. 
Actually we have always observed the bounce of the domain in a wide parameter 
range. Here, we note that we assumed $d^2x/d\tau^2(0)=0$ in these figures 
because we could not observe the drastic modification from the above 
qualitative results 
in another reasonable cases with $d^2x/d\tau^2(0) \neq 0$.
Therefore it seems that the bounce is a universal phenomena. 
Of course, the domain always collapses in the limit $\gamma \to 0$.  
Our approximation is meaningful only for $R > \langle R \rangle$ when the 
domain is well distinguished from bubbles.  
Therefore we have restricted the parameter range which respects this 
constraint.\footnote{%
We took the variable $\gamma$ as a parameter in this letter. 
In principle this is determined by the scale $\langle R \rangle$ of the bubbles 
as in \cite{sb}.  }
\par
To understand the bounce of the domain, we concentrate on the 
behavior of the solution around the radius-minimum.  
We expand the `scale factor' $a(\tau)$ 
as $a(\tau) \simeq 1+\gamma \tau x^{-2} + \cdot \cdot \cdot$, and neglect the 
velocity $\dot x(t) \approx 0$. 
Then the Euler-Lagrange equation reduces to a simple form
%============< EQUATION >==============%
%
\begin{equation}
{\ddot x}+\frac{2}{5}\lkk \frac{2x}{x^2+\gamma \tau}-\frac{1}{x} 
\rkk =:{\ddot x}+\ppp_x V(x, \tau) \simeq 0,
\end{equation}
%
%======================================%
where $V(x, \tau)=(2/5){\ln}[(x^2+\gamma \tau)/x]$. 
Explicit time dependence in this equation prevents us to use energy 
conservation law for
the analysis of kinematics of the domain.  
However, one can see rough behavior using the `potential' $V(x, \tau)$. The 
minimum of the potential is at 
$x_{\rm m}
(\tau)=(\gamma \tau)^{1/2}$ and the orbit of a domain is always 
bounded. The energy changing rate is given by 
%============< EQUATION >==============%
%
\begin{equation}
\frac{dH}{dt} \simeq \frac{2}{5}M(\langle R \rangle)
(32 \pi \langle R \rangle^3 \Gamma) \frac{\langle R \rangle}{R} \sim 
\frac{2}{5}M(\langle R \rangle) 
\frac{\frac{4 \pi}{3}\langle R \rangle^3}{\Delta V} 
(\Delta V -\Delta V')\Gamma, 
\end{equation}
%
%======================================%
where $ (\Delta V -\Delta V')\Gamma$ is the number of bubbles generated per unit 
time in 
the volume $\Delta V -\Delta V'$. The factor $(2/5)M( \langle R \rangle)$ is the 
free energy of one bubble. The factor $\frac{4 \pi}{3}\langle R \rangle^3/\Delta 
V$ is 
the minimum fraction which can contribute to the energy of the domain. Thus we 
find that the above equation is quite reasonable! 
\par
Though we have shown that a large domain always bounces and perpetually grows, 
it does not mean that the asymmetric domain eventually covers all the region of 
space. 
Actually the same argument is applicable for the domain composed of the 
symmetric phase instead of the asymmetric phase if the phase mixing is well 
established.  The effect of small bubbles are the same and they stabilize both 
types of large domains.  
This domain growth can be regarded as the percolation process that a macroscopic 
order appears and grows in the system.  
Let us clarify this point further.  
\par
We have considered the spherical symmetric domain in this letter.  This analysis 
is also applicable for the evolution of the local curvature of a boundary 
between the symmetric and asymmetric phases instead of the radius a domain.  The 
essential point in our argument was that the effect of bubble attachment is 
different in the convex side and in the concave side of the domain wall; the 
bubble attachment rate is higher in the convex side than that in the concave 
side.  Especially for a small domain, bubbles attach almost only in the outside 
and this is the reason that all the domain bounces.  
Therefore we can conclude that a large boundary wall always tend to become 
straight irrespect of the symmetry.  
This is the percolation process from the point of view of subcritical bubbles.  
We would like to report much quantitative results of percolation in our future 
publications based on the present argument.  

\vspace{1cm}

\centerline{\bf Acknowledgment}
TS thanks Humitaka \ Sato for his valuable comments. 
We would like to thank Jun'ich Yokoyama, Takahiro Tanaka 
and Masahide Yamaguchi for 
useful discussions in the initial stage of our study. 
This work is partly supported by Grant-in-Aid 
for Scientific Research Fellowship, No.\ 2925 (TS).

%\newpage

%======================================%
%<<<<<<<<<<<< REFERENCES >>>>>>>>>>>>>>%
%======================================%

\newpage
%%%%%%%%%%%%%%%%%%%%%%%%%%%%%%%%
\noindent
{\large\bf Figure Captions}   
%%%%%%%%%%%%%%%%%%%%%%%%%%%%%%%%

\noindent
Fig. 1.
\par
The variable $x \equiv R/\langle R \rangle$ vs time $\tau \equiv t/\langle R 
\rangle$.  
We fixed the initial velocity $x'(\tau)=-1$ and changed the parameter $\gamma$:  
The parameter $\gamma$ is 0.016, 0.16, 1.6, and 16 in the order of increasing 
thickness of the curve.
All domain solutions bounce and grows.  

\noindent
Fig. 2. 
\par
The same as Fig. 1, but we fixed $\gamma$ and changed the initial velocity 
$x'(\tau)=-1$ for each figures.  
$\gamma=0.16$ for Fig. 2-a,
$\gamma=1.6$ for Fig. 2-b, 
and $\gamma=16$ for Fig. 2-c.  
The parameter $x'(\tau)$ is -1, -0.5, 0, 0.5, and 1 in the order of decreasing 
length of dashing.  
%\endmulticols
\end{document}